\documentclass[twocolumn,showpacs,aps,prl,superscriptaddress,floatfix]{revtex4}

\usepackage{graphicx}
\usepackage{dcolumn}
\usepackage{amsmath}
\usepackage{epsfig}

\RequirePackage{xspace}

\usepackage{relsize}
\def\babar{\mbox{\slshape B\kern-0.1em{\smaller A}\kern-0.1em
    B\kern-0.1em{\smaller A\kern-0.2em R}}}

\def\epem       {\ensuremath{e^+e^-}\xspace}

\def\piz   {\ensuremath{\pi^0}\xspace}

\def\pip   {\ensuremath{\pi^+}\xspace}
\def\pim   {\ensuremath{\pi^-}\xspace}

\def\Kbar  {\kern 0.2em\overline{\kern -0.2em K}{}\xspace}

\def\Kz    {\ensuremath{K^0}\xspace}
\def\Kzb   {\ensuremath{\Kbar^0}\xspace}
\def\KzKzb {\ensuremath{\Kz \kern -0.16em \Kzb}\xspace}
\def\Kp    {\ensuremath{K^+}\xspace}
\def\Km    {\ensuremath{K^-}\xspace}

\def\KpKm  {\ensuremath{\Kp \kern -0.16em \Km}\xspace}
\def\KS    {\ensuremath{K^0_{\scriptscriptstyle S}}\xspace}

\def\Kstar   {\ensuremath{K^*}\xspace}

\def\Kstarm  {\ensuremath{K^{*-}}\xspace}

\def\Dbar    {\kern 0.2em\overline{\kern -0.2em D}{}\xspace}

\def\Dz      {\ensuremath{D^0}\xspace}
\def\Dzb     {\ensuremath{\Dbar^0}\xspace}
\def\DzDzb   {\ensuremath{\Dz {\kern -0.16em \Dzb}}\xspace}
\def\Dp      {\ensuremath{D^+}\xspace}
\def\Dm      {\ensuremath{D^-}\xspace}

\def\DpDm    {\ensuremath{\Dp {\kern -0.16em \Dm}}\xspace}

\def\B       {\ensuremath{B}\xspace}
\def\Bbar    {\kern 0.18em\overline{\kern -0.18em B}{}\xspace}

\def\Bz      {\ensuremath{B^0}\xspace}
\def\Bzb     {\ensuremath{\Bbar^0}\xspace}
\def\BzBzb   {\ensuremath{\Bz {\kern -0.16em \Bzb}}\xspace}
\def\Bu      {\ensuremath{B^+}\xspace}
\def\Bub     {\ensuremath{B^-}\xspace}
\def\Bp      {\ensuremath{\Bu}\xspace}
\def\Bm      {\ensuremath{\Bub}\xspace}

\def\BpBm    {\ensuremath{\Bu {\kern -0.16em \Bub}}\xspace}

\def\BorBbar    {\kern 0.18em\optbar{\kern -0.18em B}{}\xspace}
\def\DorDbar    {\kern 0.18em\optbar{\kern -0.18em D}{}\xspace}
\def\KorKbar    {\kern 0.18em\optbar{\kern -0.18em K}{}\xspace}

\mathchardef\Upsilon="7107
\def\Y#1S{\ensuremath{\Upsilon{(#1S)}}\xspace}

\mathchardef\Deltares="7101
\mathchardef\Xi="7104
\mathchardef\Lambda="7103
\mathchardef\Sigma="7106
\mathchardef\Omega="710A

\def\Deltabar{\kern 0.25em\overline{\kern -0.25em \Deltares}{}\xspace}
\def\Lbar{\kern 0.2em\overline{\kern -0.2em\Lambda\kern 0.05em}\kern-0.05em{}\xspace}
\def\Sigbar{\kern 0.2em\overline{\kern -0.2em \Sigma}{}\xspace}
\def\Xibar{\kern 0.2em\overline{\kern -0.2em \Xi}{}\xspace}
\def\Obar{\kern 0.2em\overline{\kern -0.2em \Omega}{}\xspace}
\def\Nbar{\kern 0.2em\overline{\kern -0.2em N}{}\xspace}
\def\Xb{\kern 0.2em\overline{\kern -0.2em X}{}\xspace}

\def\mes        {\mbox{$m_{\rm ES}$}\xspace}

\newcommand{\tev}{\ensuremath{\mathrm{\,Te\kern -0.1em V}}\xspace}
\newcommand{\gev}{\ensuremath{\mathrm{\,Ge\kern -0.1em V}}\xspace}
\newcommand{\mev}{\ensuremath{\mathrm{\,Me\kern -0.1em V}}\xspace}
\newcommand{\kev}{\ensuremath{\mathrm{\,ke\kern -0.1em V}}\xspace}
\newcommand{\ev}{\ensuremath{\mathrm{\,e\kern -0.1em V}}\xspace}
\newcommand{\gevc}{\ensuremath{{\mathrm{\,Ge\kern -0.1em V\!/}c}}\xspace}
\newcommand{\mevc}{\ensuremath{{\mathrm{\,Me\kern -0.1em V\!/}c}}\xspace}
\newcommand{\gevcc}{\ensuremath{{\mathrm{\,Ge\kern -0.1em V\!/}c^2}}\xspace}
\newcommand{\mevcc}{\ensuremath{{\mathrm{\,Me\kern -0.1em V\!/}c^2}}\xspace}

\def\mus  {\ensuremath{\rm \,\mus}\xspace}

\def\mus        {\ensuremath{\,\mu{\rm s}}\xspace}    

\def\to                 {\ensuremath{\rightarrow}\xspace}

\def\pep2{PEP-II}

\def\gsim{{~\raise.15em\hbox{$>$}\kern-.85em
          \lower.35em\hbox{$\sim$}~}\xspace}
\def\lsim{{~\raise.15em\hbox{$<$}\kern-.85em
          \lower.35em\hbox{$\sim$}~}\xspace}

\def\CP                {\ensuremath{C\!P}\xspace}

\xspace

\newcommand{\jprlBase}       {Phys.\ Rev.\ Lett.\xspace}

\newcommand{\jplBase}        {Phys.\ Lett.\xspace}
\newcommand{\nimBaseA}       {Nucl.\ Instr.\ Methods Phys.\ Res., Sect.\ A\xspace}

\newcommand{\nima}      [1]  {\nimBaseA~{\bf #1}}

\newcommand{\plb}       [1]  {\jplBase\ B~{\bf #1}}

\newcommand{\jprl}      [1]  {\jprlBase\ {\bf #1}}

\def\jetset74   {\mbox{\tt Jetset \hspace{-0.5em}7.\hspace{-0.2em}4}\xspace}

\def\figurebox#1#2#3{%
    \def\arg{#3}%
    \ifx\arg\empty
    {\hfill\vbox{\hsize#2\hrule\hbox to #2{\vrule\hfill\vbox to #1{\hsize#2\vfill}\vrule}\hrule}\hfill}%
    \else
    {\hfill\epsfbox{#3}\hfill}%
    \fi}

\def\rcentral{0.046}
\def\rstat{0.031}
\def\rsys{0.008}
\def\acentral{-0.22}
\def\astat{0.61}
\def\asys{0.17}
\def\asyst{0.174}

\begin{document}

\title{A Study of \boldmath{$b \to c$} and \boldmath{$b \to u$} Interference in the Decay \boldmath{$\Bm \to [\Kp\pim]_D \Kstarm$}}
%
\author{B.~Aubert}
\author{R.~Barate}
\author{D.~Boutigny}
\author{F.~Couderc}
\author{Y.~Karyotakis}
\author{J.~P.~Lees}
\author{V.~Poireau}
\author{V.~Tisserand}
\author{A.~Zghiche}
\affiliation{Laboratoire de Physique des Particules, F-74941 Annecy-le-Vieux, France }
\author{E.~Grauges}
\affiliation{IFAE, Universitat Autonoma de Barcelona, E-08193 Bellaterra, Barcelona, Spain }
% ...
\collaboration{The \babar\ Collaboration}
\noaffiliation

\begin{abstract}
Using a sample of 232$\times 10^6$ $\Upsilon(4S)\to B\bar{B}$
events collected with the \babar\  detector at the PEP-II $B$-factory
we study the decay $B^-\to [K^+\pi^-]_DK^{*-}$ where the $K^+\pi^-$ is either
from a Cabibbo-favored $\bar{D}^0$ decay or doubly-suppressed $D^0$ decay. 
We measure two observables that are sensitive to the CKM angle $\gamma$; 
the ratio $R$ of the charge-averaged branching fractions for the suppressed and favored decays; 
and the charge asymmetry $A$ of the suppressed decays:
\begin{center}
$R=\rcentral\pm\rstat(stat.)\pm\rsys(syst.)$
\end{center}
\begin{center}
$A=\acentral\pm\astat(stat.)\pm\asys(syst.).$
\end{center}
\end{abstract}

\pacs{13.25.Hw, 14.40.Nd}

\maketitle

\label{sec:Introduction}

An important feature of the standard model is that it accomodates CP violation
through the Cabibbo-Kobayashi-Maskawa (CKM) quark mixing matrix~$V$~\cite{CKM}. The self consistency of this mechanism can be
 checked by overconstraining the associated unitarity triangle~\cite{pdg}. In this paper we concentrate on the angle $\gamma$=
arg$(-V_{ud}V^*_{ub}/V_{cd}V^*_{cb})$ by studying $B$-meson decay channels where $b\to c\bar{u}s$ and
$b\to u\bar{c}s$ tree amplitudes interfere. We use a technique 
suggested by Atwood, Dunietz and Soni~\cite{ADS} (ADS) where the final state
$B^-\to K^+\pi^- K^*(892)^-$  can be reached from two amplitudes, $B^-\to D^0 K^{*-}$ followed by the 
doubly-Cabibbo-suppressed decay $D^0\to K^+\pi^-$, and 
$B^-\to \bar{D}^0 K^{*-}$ followed by the Cabibbo-favored decay $\bar{D}^0\to K^+\pi^-$~\cite{chargeconj}. 
The size of the interference between these two amplitudes 
depends on the CKM angle $\gamma$ as well as the \CP-conserving
relative strong phases $\delta_B$ and $-\delta_D$, and the ratios $r_B$ and $r_D$
of suppressed and favored amplitude magnitudes in \B-
($\mathcal{A}(B^-\to \bar{D}^0 K^{*-})$ and $\mathcal{A}(B^-\to D^0 K^{*-})$), and 
$D$- ($\mathcal{A}(D^0\to K^+\pi^-)$ and $\mathcal{A}(D^0\to K^-\pi^+)$) decays.
We define two measurable quantities, $R$ and $A$, as follows:
\begin{eqnarray}
R&=&\frac{\Gamma(B^-\to[K^+\pi^-]_DK^{*-})+\Gamma(B^+\to[K^-\pi^+]_DK^{*+})}
{\Gamma(B^-\to [K^-\pi^+]_DK^{*-})+\Gamma(B^+\to [K^+\pi^-]_DK^{*+})}, \nonumber \\
A&=&\frac{\Gamma(B^-\to[K^+\pi^-]_DK^{*-})-\Gamma(B^+\to[K^-\pi^+]_DK^{*+})}
{\Gamma(B^-\to[K^+\pi^-]_DK^{*-})+\Gamma(B^+\to[K^-\pi^+]_DK^{*+})}. \nonumber
\end{eqnarray}
\noindent
The notation $[K^+\pi^-]_D$ indicates that these particles are neutral
$D$-meson ($\Dz or \Dzb$) decay products. Neglecting the very small effect
of $D^0\bar{D}^0$ mixing as justified in~ref.{\cite{ADS}, 
$R$ and $A$ are related to $\gamma$, the strong phases, $r_B$, and $r_D$ by
\begin{eqnarray}
R&=&r_D^2+r_B^2+2r_Dr_B\cos(\delta_B+\delta_D) \cos\gamma,
\label{eq:R} \\
A&=&2r_Dr_B\sin(\delta_B+\delta_D)\sin\gamma/R.
\label{eq:A}
\end{eqnarray}
\noindent
In the above equations only $r_D$ has been measured: $r^2_D=0.00362\pm 0.00029$~\cite{pdg}.
Estimates for $r_B$ are in the range $0.1\le r_B\le 0.3$~\cite{rb}.
Because there are more unknowns than measurable quantities, determining $R$ and $A$ does not
uniquely determine $\gamma$. However, the $R$ and $A$ measured here can be used in combination with
a similar technique proposed by Gronau, London, and Wyler (GLW)~\cite{GLW} to provide 
constraints on $r_B$ and eventually $\gamma$. Other methods sensitive to
$\gamma$~\cite{ADS,Giri:2003ty}  rely on the analysis of three-body \Dz final states.

\label{sec:babar}

This analysis uses data collected near the $\Upsilon(4S)$ resonance with the \babar\  detector~\cite{babarD}
at the PEP-II storage ring. The data set consists of 211 fb$^{-1}$ collected at the peak of the $\Upsilon(4S)$
(232$\times 10^6~B\bar{B}$ pairs) and 20.4~fb$^{-1}$ 40~\mev below the resonance peak~(off-peak data). 

The \babar\  detector uses a five-layer double-sided silicon vertex tracker (SVT) and a 
40-layer drift chamber (DCH) to measure the trajectories of charged particles. Both the SVT and
DCH are located inside a 1.5-T solenoidal magnetic field. Photons are detected by means of a CsI(Tl)
crystal calorimeter also located inside the magnet. Charged particle
identification is determined from information provided by a ring-imaging Cherenkov device (DIRC) in combination with ionization
measurements ($dE/dx$) from the tracking detectors. 
The \babar\  detector's response to various physics processes as well
as varying beam and environmental conditions is modeled with GEANT4~\cite{geant4} based software.    

\label{sec:Analysis}

The decay $B^-\to D^0K^{*-}$ is reconstructed in final states where the $K^{*-}$
decays to $\KS\pi^-$ followed by $\KS\to \pi^+\pi^-$ and the $D^0$ decays into a
charged kaon and pion. The analysis begins with the selection of $\KS$ candidates from 
oppositely charged tracks assumed to be pions.
The invariant mass of the $\KS$ candidate is required to be within
10 $\mevcc$ (about three standard deviations) of the 
nominal $\KS$ mass~\cite{pdg}.
The $\KS$ candidate is required to travel at least four times further
than the standard deviation of its decay length. 
Its flight direction and decay length must be consistent with those of a $\KS$ 
originating from the interaction point. The momentum of a $\KS$ candidate meeting these criteria 
is then recalculated with a mass and vertex constraint. Next, a $\KS$ is paired with a charged track, assumed
to be a pion, and the combination is constrained to come from the interaction point.
The pair is kept for further study if its invariant mass is within 55 $\mevcc$ of the
nominal $K^*$ mass~\cite{pdg}. Finally, since the $K^*$ from a $B^-\to D^0 K^{*-}$ decay is polarized,
we require $|\cos\theta_H| \ge 0.4$ where $\theta_H$ is the angle in the $K^*$ rest frame between the daughter
pion and the parent $B$ momentum vector. This helicity-angle
requirement helps discriminate \B mesons from combinatorial background (mostly $e^+e^-\to q\bar{q}$ continuum events;
$q\in \{u,d,s,c\}$) as the former have a $\cos^2\theta_H$ distribution
while that of the latter is uniform.  

To perform the measurement, we reconstruct Cabibbo-favored $\Dz \to\Km\pip$ and doubly-Cabibbo-suppressed $\Dz \to \Kp \pim$ 
candidates. Candidates that have an invariant mass within
18 $\mevcc$ (2.5 standard deviations) of the nominal $D^0$ mass~\cite{pdg} are kept for further study.
We also select $\Dz \to K^-\pi^+\pi^0$ and $\Dz \to K^-\pi^+\pi^+\pi^-$ 
candidates to define various signal distributions discussed later in this paper. 
Loose particle identification criteria are imposed on the charged
particles of all studied decay channels. 
Pairs of photons with a total energy greater than 200~\mev and an
invariant mass in the range $125\le m_{\gamma\gamma}\le 145$ $\mevcc$
are combined to form $\piz$ candidates that are refit, with their
invariant mass constrained to the nominal \piz mass~\cite{pdg}. Loose
kinematic criteria are used to select the three- and four-body candidates.   

Suppression of backgrounds from $e^+e^-\to q\bar{q}$ continuum events 
is achieved by using 
event shape and angular variables. 
Global event-shape
variables are used to eliminate events with jet-like topology, a
signature of $e^+e^-\to q\bar{q}$ continuum events.
The thrust angle of a $B$-meson candidate is required to satisfy $|\cos\theta_T| \le 0.9$, where
$\theta_T$ is the angle between the thrust axis of the $B$-meson and that of the rest of the event.    

To further reduce the $q\bar{q}$ contribution to our data sample a
neural network (NN) is used.
The variables used in the neural network consist of the 
angular moments $L^0$ and $L^2$ defined in Ref.~~\cite{bib:muriel}, the ratio $R_2=H_2/H_0$ of
Fox-Wolfram moments~\cite{foxwolfram}, 
the $\chi^2$ of the \B-meson vertex fit, the cosine of the angle between
the $B$ candidate momentum vector and the beam axis ($\cos\theta_B$),
$\cos\theta_T$ (defined above), and the cosine of the
angle between a $D^0$ daughter momentum vector in the $D^0$
rest frame and the direction of the $D^0$ in the \B-meson rest frame ($\cos\theta_H(D^0)$).
The NN is trained with signal Monte Carlo events and continuum data
collected below the $\Upsilon(4S)$ (off-peak data).
The NN is then cross-checked with an independent set of signal Monte
Carlo events. Finally, we verify that the NN has a consistent output for off-peak data
and $q\bar{q}$ Monte Carlo events. 
The separation between signal and continuum background is shown in Fig.~\ref{fig:nnPerformance}. 
We select candidates with neural network output above 0.8. 
Our event selection is optimized to minimize the statistical error on the signal yield, 
determined using simulated signal and background events.
\begin{figure}[ht]
\begin{center}
\includegraphics[width=9cm]{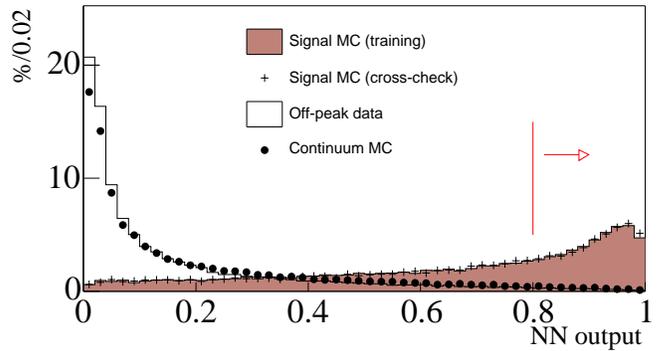}
\caption{\label{fig:nnPerformance}
The result of the neural network training and verification (see text). 
The training samples are shown as histograms. 
The signal (Monte Carlo simulation) is the shaded histogram peaking to
the right; the background (off-peak data recorded
40 \mev below the resonance) is the histogram with a peak near~0.
The data samples used to check the NN are overlaid as data points.The
vertical bar and the arrow indicate the requirement used to select
signal candidates.}
\end{center}
\end{figure}

We identify $B$-meson candidates using two nearly independent variables that
take advantage of the well defined beam energy and the known kinematics of $\Upsilon(4S)$ decay:
the beam-energy-substituted mass $m_{ES}=\sqrt{(s/2+{\bf p}_0\cdot {\bf p}_B)^2/E^2_0-p^2_B}$
and the energy difference $\Delta E=E^*_B-\sqrt{s}/2$ where the subscripts 0 and $B$ refer to
the $e^+e^-$ system and $B$-meson candidate, respectively; $\sqrt{s}$ is the $\epem$
center-of-mass (CM) energy and the asterisk labels the CM frame. The $m_{ES}$ distribution for signal events is
well
represented by a Gaussian function with mean centered at the known mass of the $B^-$~\cite{pdg}
and width 2.76 $\mevcc$. 
The $\Delta E$
distribution for signal events is described by a Gaussian function centered at zero with a width
that varies from 11 to 13 $\mev$ among the different final states. These quantities are measured in the
data from $\Bm \to \Dz \Kstarm$ with Cabibbo-favored $\Dz$ decays. For this analysis signal events must
 satisfy $|\Delta E|\le 25$ \mev.  

The efficiency to detect a $\Bm \to \Dz \Kstarm$ signal event where
$\Dz \to K \pi$, after all criteria are imposed, is (9.6$\pm0.1$)\%. This efficiency is 
the same for $D^0\to K^-\pi^+$ and $D^0\to K^+\pi^-$. There are
multiple candidates in  12\% of the events. In such cases 
the candidate with the smallest $|\Delta E|$ is selected for further
study. According to Monte Carlo simulation, this is the correct candidate 88\% of
the time.

We study various potential sources of background using a combination
of Monte Carlo simulation and data events. Two
sources of background are identified in large samples of simulated $B\bar{B}$ events. One source
is $\Dz\KS\pim$ production 
where the $\KS\pi^-$ is non resonant and has an invariant mass in the $K^{*-}$ mass window. This background is discussed
later in this paper. The second background (peaking background) includes instances where a favored decay
(i.e. $B^-\to [K^-\pi^+]_DK^{*-}$) contributes to fake candidates for the suppressed decay (i.e., $B^+\to [K^-\pi^+]_DK^{*+}$).
The most common way for this to occur is for a $\pi^+$ from the rest of the event to be substituted for
the $\pi^-$ in the $K^{*-}$ candidate. 
Other sources of peaking background include double particle-identification failure in signal events
that results in $D^0\to K^-\pi^+$ being reconstructed as $D^0\to \pi^-  K^+$, or
the kaon from the $D^0$ being interchanged with the charged pion from the $K^*$. 
From a detailed Monte Carlo study the total size of this background is 
estimated to be 1.4$\pm$0.2 events. 
We also verify with the Monte Carlo simulation that the charmless decays with the
same final state as the signal (e.g., $B^-\to K^{*-}K^-\pi^+$) 
are not a significant background for this analysis.

Signal yields are determined from an unbinned extended maximum likelihood fit to the $m_{ES}$ distribution
in the range 
$m_{ES}\ge 5.2\ \gevcc$. A Gaussian function ($\mathcal{G}$) is used to describe all signal
shapes while the combinatorial background is modeled with an ARGUS~\cite{argus} threshold function($\mathcal{A}$).     
This function's shape is determined by one parameter $\xi$ while a
second parameter, E$_{\rm max}=\sqrt{s}/2$,
(fixed at $5.2901\ \gevcc$) is the maximum mass for pair-produced
\B-mesons given the collider beams energies. For a probability distribution function ($PDF$) we use
$a\cdot \mathcal{A}+b\cdot \mathcal{G}$ where $a$ is the number of background events and $b$ the number of signal events.
We correct $b$ for the peaking background previously discussed (1.4$\pm$0.2 events).
The mean and width of $\mathcal{G}$ and the value of $\xi$ are
determined by an initial fit to all $B^-\to D^0K^{*-}$ candidates
where the $D^0$ decays into the Cabibbo-favored channels $K^-\pi^+,~K^-\pi^+\pi^0$, and $K^-\pi^+\pi^+\pi^-$.
\begin{figure}[htb]
\begin{center}
\includegraphics[width=6cm]{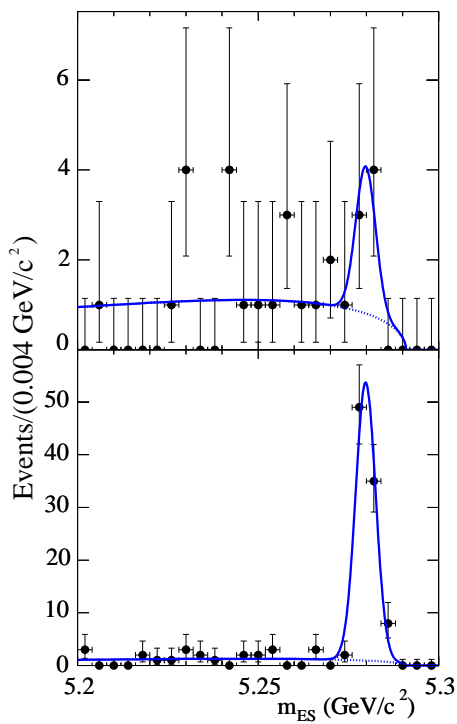}
  \caption{\label{fig:adsfit}
  Distributions of \mes for the wrong-sign (top) and right-sign
  (bottom) decays. These decay categories are defined in the text. 
  The curves result from a simultaneous fit to
  these distributions with identical PDFs for both samples.}
\end{center}
\end{figure}
\begin{figure}[hbt]
\begin{center}
\includegraphics[width=6cm]{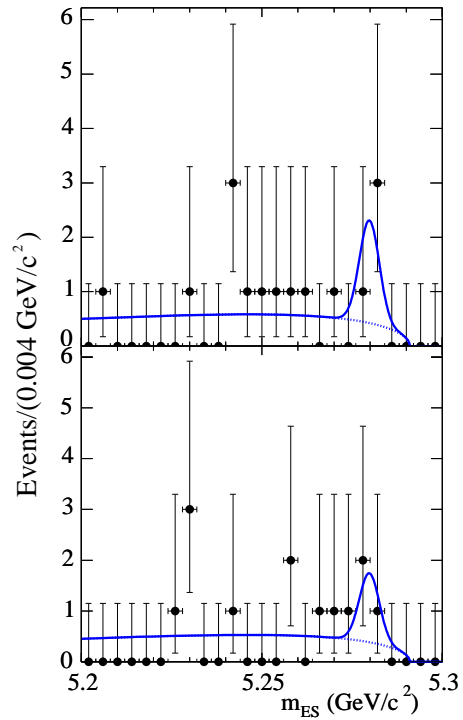}
  \caption{\label{fig:adsfitA}
Result of a simultaneous fit to the wrong-sign  and right-sign \mes distributions
with identical PDFs for both samples. Here the wrong-sign sample shown
in the top plot of Fig.~\ref{fig:adsfit} is split by charge to measure $A$. 
The upper plot shows the \mes distribution of the $B^+\to
[K^-\pi^+]_DK^{*+}$ decays 
while the lower plot presents the same for the $B^-\to [K^+\pi^-]_DK^{*-}$
decays. The curves are the results of the fit.}
\end{center}
\end{figure}

In Fig.~\ref{fig:adsfit} we show the results of a simultaneous fit to $B^-\to [K^+\pi^-]_DK^{*-}$ and
$B^-\to [K^-\pi^+]_DK^{*-}$ candidates that satisfy all selection
criteria. We call wrong- (right-) sign decays those where the $\Kstar$ and the
kaon have opposite (same) strangeness. It is in the wrong-sign decays that the interference
we study takes place.  Therefore in Fig.~\ref{fig:adsfitA}
we display the same fit separately for the wrong-sign decays of the
\Bp and the \Bm mesons. 
The results of the maximum likelihood fit are $R=0.046\pm0.031$, $A=-0.22\pm0.61$, and $91.2\pm9.7 \ B^-\to
[K^-\pi^+]_DK^{*-}$ right-sign events. Expressed in terms of the wrong-sign yield, the fit result is
$4.2 \pm 2.8$ wrong-sign events. The errors are statistical only. The
correlation between $R$ and $A$ is insignificant. 

In Table~\ref{tab:ADS} we summarize the systematic errors relevant to this analysis. 
Since both $R$ and $A$ are ratios of similar quantities most potential sources of systematic errors cancel.
The estimate for the detection-efficiency asymmetry is obtained from a sample of $B^-\to D^0\pi^-$ events.
Here a charge asymmetry of $A_{ch}=(-1.9\pm0.8)$\% is measured. We add linearly
the central value and one-standard deviation in the most
conservative direction to assign a systematic error of 
$\delta A_{ch}=\pm0.027$ to the $A$ measurement. To a good approximation
the systematic error in $R$ due to this source can be shown to be given by $\delta R=R\cdot A\cdot \delta A_{ch}$, with
$A$ the previously determined central value.

\label{sec:Systematics}

To estimate the systematic uncertainty on $A$ and $R$ due to the peaking background, we use the statistical
uncertainty on this quantity, $\pm0.2$ events. With approximately $4\ B^-\to [K^+\pi^-]_DK^{*-}$ events
and $90\ B^-\to [K^-\pi^+]_DK^{*-}$ events this source contributes $\pm 0.002$ and 
 $\pm0.043$ to the systematic errors on $R$ and $A$, respectively.

To account for the non resonant $\KS\pi^-$ pairs in the $K^*$ mass range 
we study a model that incorporates  S-wave pairs in both the $b\to c\bar{u}s$ and $b\to u\bar{c}s$
amplitudes. It is expected that higher order partial waves do not
contribute. The amount of S-wave present in the favored $b\to c\bar{u}s$ amplitude is 
determined directly from the data by examining the angular
distribution of the $\KS\pi$ system in the $\Kstar$ mass region. To estimate
the systematic errors 
due to this source we vary all strong phases
within 2$\pi$ and calculate the maximum deviation between the S-wave model and the expectation
if there were no non resonant contribution for both $R$ (Eq.~\ref{eq:R}) and $A$ (Eq.~\ref{eq:A}).
The result of this study is given in Table~\ref{tab:ADS}.

In the fit to the $m_{ES}$ distribution we use parameters determined from an initial fit that
includes $D^0\to K^-\pi^+,~K^-\pi^+\pi^0$ and $K^-\pi^+\pi^+\pi^-$
candidates. We repeat the fit
using parameters from each individual $D^0$ decay channel and calculate a contribution to the
systematic error from the observed variations in $R$ and $A$. We also
vary E$_{\rm max}$ by $\pm 2\mev$, refit and add the maximum shifts
in $R$ and $A$ to the previously described contribution to obtain the 
systematic error associated with the shape of the \mes distribution. 
After adding in quadrature the  individual systematic error contributions listed in
Table~\ref{tab:ADS}, we find:
\begin{eqnarray}
R&=&\rcentral\pm\rstat(stat.)\pm\rsys(syst.), \nonumber \\
A&=&\acentral\pm\astat(stat.)\pm\asys(syst.). \nonumber 
\end{eqnarray}
We also quote the results in terms
of two other variables:
\begin{eqnarray}
R\ (1 + A)&=&0.036 \pm 0.042 \pm 0.010, \nonumber \\
R\ (1 - A)&=&0.056 \pm 0.045 \pm 0.012. \nonumber 
\end{eqnarray}
They may be of use in combining $\gamma$-sensitive measurements from the GLW and ADS
methods, and the analyses exploiting three-body \Dz decays. 
The effect of the non resonant $\KS\pim$ background gives the
dominant contribution to the systematic uncertainties,~$\pm 0.009$ on
both quantities.
\begin{table}[h]
\begin{center}
\caption{\label{tab:ADS}
Summary of systematic uncertainties.
}
\begin{tabular}{lcc}
\hline \hline
Source                              & $\delta R$            & $\delta A$\\
\hline
Detection asymmetry                 &$\pm0.0003$            & $\pm0.027$\\
Peaking background                  &$\pm0.002\ $           & $\pm0.043$\\
Non resonant $K_s^0\pi^-$ background&$\pm0.0073$            & $\pm0.126$\\
Shape of $m_{ES}$ distribution      &$\pm0.0023$            & $\pm0.108$\\
\hline
Total systematic error              &$\pm\rsys\ $           &$\pm\asyst$\\
\hline \hline
\end{tabular}
\end{center}
\end{table}
\label{sec:Physics}

\begin{figure}[ht]
\begin{center}
\includegraphics[width=8cm]{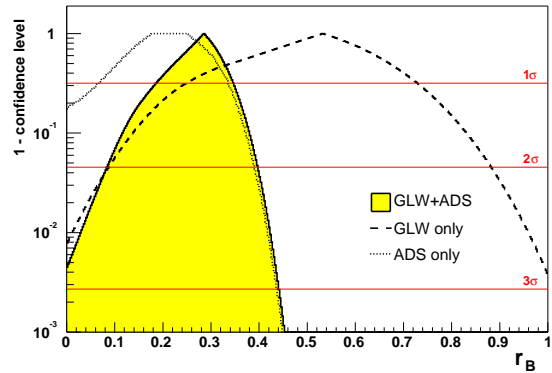} 
 \caption{\label{fig:rb_log}
 Constraints on $r_B$. The \babar\  $\Bm\to D_{CP}K^{*-}$
 (GLW)~\cite{babar_glw} result is combined with this
analysis. The dashed (dotted) curve shows 1 minus the
confidence level to exclude the abscissa-value as a function of $r_B$ derived from the GLW
(ADS) only measurements. When both the GLW and ADS results are
combined the curve above the shaded area is obtained. Horizontal
lines show the exclusion limits at the 1, 2 and 3 standard deviation levels.}
\end{center}
\end{figure}
\begin{figure}[hbt]
\begin{center}
\includegraphics[width=9cm]{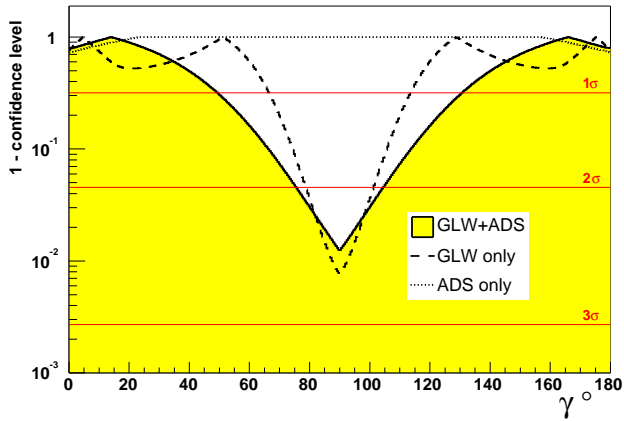} 
  \caption{\label{fig:gamma_log}
 1 minus exclusion confidence level curve for $\gamma$ obtained from the \babar\  
$\Bm\to D_{CP}K^{*-}$ (GLW) result combined with this
analysis. The graphical conventions are described in the caption of Fig.~\ref{fig:rb_log}.}
\end{center}
\end{figure}

In order to extract information on $r_B$ and $\gamma$ we combine the above measurements of $R$ and
$A$ with measurements of similar quantities, $R_{CP\pm},~A_{CP\pm}$,  from
$B\to D^0_{CP}K^{*-}$ ~\cite{babar_glw} using the method suggested
in Ref.~\cite{GLW}, in which the \Dz decays to \CP eigenstates are exploited. 
A frequentist statistical approach~\cite{ckmfitter} is used. A $\chi^2$ is formed
from the differences between the measured and theoretical values, and the covariance matrix
of the six measured variables. We restrict $r_B$ to values between 0 and 1.3 and allow $\gamma$
to vary between 0 and 180$^{\circ}$ for all possible values of $(\delta_B+\delta_D)$ between 0 and 360$^{\circ}$. 
We call $\chi^2_{min}(=1.4)$ the minimum $\chi^2$ for the whole parameter space. We then scan the $r_B$ range: 
for each value of $r_B$ we minimize the $\chi^2$ across the reduced parameter space (where $r_B$ is fixed), and
find  $\chi^2_m$. We use $\Delta \chi^2 = \chi^2_m - \chi^2_{min}$ to
compute the confidence level of $r_B$ assuming Gaussian uncertainties. 
Figure~\ref{fig:rb_log} shows the confidence level resulting from this $r_B$ scan. 
Combining the ADS and GLW results we find
$$r_B=0.28^{+0.06}_{-0.10}.$$
\noindent
In a similar fashion, we show the confidence level for the $\gamma$ scan in Fig.~\ref{fig:gamma_log}.
The interval $75^{\circ} \le \gamma \le 105^{\circ}$ is disfavored at the two-standard deviation level.

In summary we present the first measurements of yields from $B^-\to
[K^+\pi^-]_D K^{*-}$ decays.  
By exploring the behavior of the likelihood function close to its
maximum, we determine that the  statistical significance for $R$ to
differ from zero is at the two-standard deviation level. As seen on
Fig.~\ref{fig:rb_log}, this (ADS) result narrows the allowed $r_B$
range previously obtained with the GLW method~\cite{babar_glw}.
The constraint the ADS method provides on $\gamma$ is weak with the present data sample.

We are grateful for the excellent luminosity and machine conditions
provided by our \pep2\ colleagues, 
and for the substantial dedicated effort from
the computing organizations that support \babar.
The collaborating institutions wish to thank 
SLAC for its support and kind hospitality. 
This work is supported by
DOE
and NSF (USA),
NSERC (Canada),
IHEP (China),
CEA and
CNRS-IN2P3
(France),
BMBF and DFG
(Germany),
INFN (Italy),
FOM (The Netherlands),
NFR (Norway),
MIST (Russia), and
PPARC (United Kingdom). 
Individuals have received support from CONACyT (Mexico), A.~P.~Sloan Foundation, 
Research Corporation,
and Alexander von Humboldt Foundation.


\begin{thebibliography}{99}
\bibitem{chargeconj}
Charge conjugate decays are implied throughout this paper unless
explicitly otherwise stated.
\end{thebibliography}
\end{document}